\documentclass[manuscript]{aastex}

\slugcomment{Submitted, Version 4.0, 140815}

\shorttitle{A Closer Look at the 2012-B eruption of SN 2009ip}
\shortauthors{Martin et al.}

\begin{document}


\title{A Closer Look at the Fluctuations in Brightness of SN 2009ip During Its Late 2012 Eruption}


\author{J. C. Martin}
\affil{Barber Observatory, University of Illinois Springfield, Springfield, IL 62704, USA}
\email{jmart5@uis.edu}

\author{F.-J. Hambsch}
\affil{Remote Observatory Atacama Desert, Chile\\
Vereniging Voor Sterrenkunde (VVS), Oude Bleken 12, 2400 Mol, Belgium}

\author{R. Margutti}
\affil{Harvard-Smithsonian Center for Astrophysics, 60 Garden St, Cambridge, MA  02318, USA}

\author{T.G. Tan}
\affil{Perth Exoplanet Survey Telescope, Perth, Australia}

\author{I. Curtis}
\affil{Adelaide, Australia}

\and

\author{A. Soderberg}
\affil{Harvard-Smithsonian Center for Astrophysics, 60 Garden St, Cambridge, MA  02318, USA}



\begin{abstract}
The supernova impostor SN 2009ip has re-brightened several times since its initial discovery in August 2009.  During its last outburst in late September 2012 it reached a peak brightness of m$_v$ $\sim$ 13.5 (M$_v$ brighter than -18) causing some to speculate that it had undergone a terminal core-collapse supernova.   Relatively high-cadence multi-wavelength photometry of the post-peak decline revealed bumps in brightness infrequently observed in other Type IIn supernovae.  These bumps occurred synchronously in all UV and optical bands with amplitudes of 0.1 -- 0.4 mag at intervals of 10 -- 30 days.  Episodic continuum brightening and dimming in the UV and optical  with these characteristics is not easily explained within the context of models that have been proposed for the late September 2012 outburst of SN 2009ip.   We also present evidence that the post peak fluctuations in brightness occur at regular intervals and raise more questions about their origin.

\end{abstract}




\section{Introduction\label{intro}}

\cite{2009CBET.1928....1M} first identified SN 2009ip in August 2009 as an under-luminous supernova in NGC 7259 (z $\sim$ 0.005, d $\sim$ 24 Mpc).    Based on further investigation and subsequent eruptions in 2010 and 2011, it was reclassified as a supernova impostor.  For additional observational details concerning observed activity of SN 2009ip prior to 2012 see \cite{2009ATel.2184....1B}, \cite{2009ATel.2212....1L}, \cite{2010AJ....139.1451S}, \cite{2011ApJ...732...32F}, \cite{2013ApJ...767....1P}.

There were two significant outbursts in 2012: 2012-A and 2012-B.  The first on UTC 2012 July 24  (2012-A) \citep{2012ATel.4334....1D} reached a peak of m$_V$ $\sim$ 16.8 and faded. Then a UTC 2012 September 15 spectrum by \cite{2012ATel.4412....1S} showed signs of a second outburst (2012-B), which they announced as a Type IIn core-collapse supernova.  Observations on UTC 2012 September 22 \citep{2012ATel.4414....1M} and UTC 2012 September 24 \citep{2012ATel.4416....1M} detected no change in brightness (m$_V$ $\sim$ 17.7).  But soon after that, SN 2009ip brightened more than three magnitudes in less than 50 hours \citep{2012ATel.4423....1B,2012ATel.4425....1M,2013ApJ...763L..27P}.  

The 2012-B outburst was the brightest observed yet for SN 2009ip, reaching a peak brightness of m$_V$ $\sim$ 13.5 (M$_V$ $<$ -18).  Consequently, several authors have analyzed the 2012-B outburst and considered evidence that it was a terminal core-collapse supernova explosion.  Some have concluded that it was a terminal explosion   \citep{2013ApJ...763L..27P,2013MNRAS.430.1801M,2013arXiv1308.0112S}.   Others have questioned if it was a core-collapse supernova \citep{2013ApJ...767....1P,2013MNRAS.433.1312F,2013ApJ...764L...6S,2013arXiv1306.0038M}.   

\cite{2013MNRAS.433.1312F}, \cite{2013arXiv1306.0038M} and \cite{2014ApJ...787..163G} noted brightness fluctuations in the decline of the 2012-B outburst that are not present in most other well sampled Type IIn light curves \citep{2012ApJ...744...10K,2013A&A...555A..10T}.  Similar fluctuations maybe present in a subset of Type IIn light curves including SN 1994W \citep{1998ApJ...493..933S}, SN 2005la \citep{2008MNRAS.389..131P}, SN 2005ip \citep{2009ApJ...695.1334S} and SN 2010mc \citep{2013Natur.494...65O} but they have not received detailed attention.

The observed bumps in brightness were a change of several tens of percent or more relative to the total flux.  To our knowledge, there is no record of comparable variations in flux of the major features in the emission spectrum that can to an order of magnitude account for these bumps  \citep{2013MNRAS.430.1801M,2013MNRAS.433.1312F,2013arXiv1306.0038M}.  We have not conducted an independent search of the published spectra and the sampling may be too sparse to detect a clear correlation between relatively small changes in the emission features and many of the bumps.  \cite{2014ApJ...787..163G}  identified a modest increase in H-alpha equivalent width that coincides with the bump in brightness 20 -- 40 days after the explosion.  that is explained as rise and decline in the continuum rather than a change in the absolute flux of the emission feature.  While there is no obvious correlation with large scale changes in the flux of the emission features, the bumps are clearly present in the derived temperature and bolometric luminosity of blackbody continuum fits to the photometry \citep{2013arXiv1306.0038M}.  \cite{2014ApJ...787..163G} confirms that several of the bumps correlate with blueward deviations in the observed color.  This may imply that the bumps are primarily driven by changes in the continuum brightness.

Bright emission lines in the spectra of Type IIn supernovae may vary as the post-break-out shock interacts with density variations in the circumstellar medium \citep{1994MNRAS.268..173C,2002ApJ...572..350F}.  These occur in an optically thin regime that almost exclusively produces radio and X-ray emission, without significant visual continuum brightening \citep{1994ApJ...420..268C,2006A&A...449..171N,2012MNRAS.419.1515D,2012ApJ...755..110C}.   When the supernova is embedded in an optically thick envelope, the radiation produced by the shock is processed before reaching the observer  \citep{1977ApJS...33..515F,2011ApJ...729L...6C,2011MNRAS.414.1715B}. Under these circumstances, high energy photons produced in shock interactions may emerge as optical continuum after passing through material of sufficient optical depth \citep{2011PhRvD..84d3003M,2012ApJ...747L..17C,2012ApJ...759..108S}.    

 The most basic models assume that energy uniformly diffuses outward through successive layers until it streams freely through space to the observer.  In reality, the structure of the circumstellar environment may contain asymmetric clumps or large structures that break spherical or bipolar symmetry.   Spectra and polarimetry provide evidence that both 2012 eruptions of SN 2009ip produced aspheric outflows \citep{2014MNRAS.442.1166M}.
Computer simulations show that geometry, density gradient, and other qualities of the circumstellar envelope may significantly influence the observed bolometric light curve for the supernova \citep{2010MNRAS.407.2305V,2007Natur.450..390W,2007ApJ...671L..17S,2013MNRAS.435.1520M}.   However it is still uncertain how asymmetric or more complicated geometry may affect the observed light curve and spectrum.  

\cite{2013MNRAS.430.1801M} propose a scenario for SN 2009ip where the 2012-A event was a terminal core-collapse and the interaction of that shock wave with an already existing dense circumstellar envelope produced the 2012-B peak brightness.  
They conclude based on spectra of high velocity ($>$ 8000 km s$^-1$) ejecta that SN 2009ip underwent a terminal core-collapse supernova explosion at the time of the 2012-A outburst.   \cite{2013ApJ...763L..27P} agree but \cite{2008MNRAS.389..131P} question if high velocity ejecta alone is ``conclusive proof'' of a terminal core-collapse.  \cite{2013arXiv1306.0038M} demonstrated that the expansion of the photosphere during the 2012-B event is too fast to have originated in the 2012-A event and calculate the total radiated energy of the 2012-B eruption to be 3.2$\times$10$^{49}$ ergs.   
 Both \cite{2013MNRAS.433.1312F} and \cite{2013arXiv1306.0038M} also found {\em{no}\em} evidence of supernova nucleosynthesis in late-time spectra (t $>$ +150 days).   \cite{2013arXiv1308.0112S} argue that the radiated energy could account for 5\% or less of the total energy  and speculate about alternatives for the absence of oxygen rich material in the late-time spectra.  \cite{2014ApJ...787..163G} state spectra taken more than 70-days after peak brightness are more consistent with a Type IIn SNe than an LBV-like outburst but offer the possibility that not all Type IIn SNe are terminal explosions.

The term ``photosphere'' can be confusing because it commonly evokes a uniform opaque surface or shell completely enclosing a source of energy diffusing through it.   Following the lead of \cite{2013arXiv1306.0038M} we attempt to avoid this confusion by referring to the source of optical and UV continuum photons in the SN 2009ip spectrum as the ``continuum emitting region'' independent of model, geometry or source of heating.

Considering alternative scenarios, the bumps in the light curve could be caused by heating \em{in addition }\em to the main shock-interaction.  The additional heating may come from a surviving progenitor, neutron star or accretion disk around a black hole produced in a core-collapse supernova.  Or one referee suggested the opposite:  that embedded source heating could be the dominant mechanism for the late-time light curve and the fluctuations themselves are powered by the addition of power from interaction with an inhomogeneous CSM. \cite{2014ApJ...787..163G} consider a scenario along these lines.  They identify some correlation between the timing of the bumps in brightness relative to past activity of SN 2009ip.  A detailed computational analysis of these alternatives is outside the scope of this paper.

Slow moving ejecta may  also interact with the circumstellar medium at later times (after peak brightness) adding to the observed brightness.  \cite{2012arXiv1211.4577L} postulate that the 2012-B brightening was due to the illumination of the inner rim of a dense equatorial disk through interaction with fast moving ejecta.  They argue that the Balmer line ratios imply a high density over a relatively small surface area so that a disk geometry is favored over shock wave interaction with spherical shells.  This presents an interesting alternative that could be consistent with the bumps in the decline.  The continuum emitting region situated in the disk could be reheated in episodes throughout the decline when slow moving or later ejected fast moving material interacts with the inner rim {\em{after}\em} the initial shock wave passed.  \cite{2013arXiv1308.0112S} find no spectroscopic evidence of slower moving ejecta interacting at later times but this does not rule out faster ejecta originating from the source after the peak brightness.

\cite{2013ApJ...764L...6S} and \cite{kashi} provide a model for the 2012-B eruption  involving a “merger burst” in a binary star system.  They focus on the two largest fluctuations in the decline at t= +30 and t= +60 days.  In their model those peaks mark interactions between stars in a binary system during periastron passage.  The excreted material impacts slower moving optically thick material, heating it and producing observable continuum.  Material excreted from the binary continues to impact and heat the continuum emitting region throughout the decline at regular intervals associated with the frequency of the binary orbit.

This paper extracts and attempts to characterize the brightness fluctuations in the light curve.  It is organized into three sections.  In Section \ref{obs} we describe the photometry and how it was measured.  In Section \ref{analysis} we analyze the data and characterize the fluctuations.  A discussion and conclusions are provided in Section \ref{discuss}.

All uncertainties given in this paper are one-sigma unless otherwise noted.  We use U, B, V, Rc, and Ic to refer to standard Johnson-Cousins filters.  The Swift/UVOT filters are referred to with lowercase b, u, w1, m1, and w2.  Magnitudes are zero-pointed to Vega on the Johnson System.  Time is referred to in the manner adopted by \cite{2013arXiv1306.0038M} where “t” is given in days and t = 0 is MJD 56203 (October 3, 2012 UTC) corresponding with the peak UV brightness.  Where observed flux is converted to luminosity we make the same assumptions as \cite{2013arXiv1306.0038M} and \cite{2011ApJ...732...32F}:  a distance of 24 Mpc and a Milky Way extinction $E(B-V) = 0.019$ \citep{1998ApJ...500..525S} with no extinction in the host galaxy.

\section{Observations\label{obs}}

\subsection{Ground-Based Optical Measurements}
Images of SN 2009ip were obtained in the V, Rc and Ic bands using four telescopes:  the University of Illinois Springfield (UIS) Barber Observatory 20-inch telescope with an Apogee U42 CCD camera using a back-illuminated E2V CCD42-40 chip (Pleasant Plains, IL), a 16-inch f/6.8 reflecting telescope with a Finger Lakes Instruments (FLI) CCD camera using a Kodak 16803 chip operated by Franz-Josef Hambsch at the Remote Observatory Atacama Desert \citep{2012JAVSO..40.1003H}, the Perth Exoplanet Survey Telescope (PEST)\footnote{http://www.angelfire.com/space2/tgtan/PEST\_description.pdf} 
a 12-inch Meade LX200 schmidt-cassegrain telescope (SCT) with an SBIG ST-8MXE CCD camera using a Kodak 0400 chip operated by TG Tan (Perth, Australia), and a C11 SCT with a ATIK 320E mono CCD camera using a Sony ICX274 chip operated by Ivan Curtis (Adelaide, Australia).  The Barber Observatory imaged all bands with 600-second exposures.  Hambsch obtained 240-second exposures in V and Ic bands in pairs.  

Curtis and Tan took a series of high cadence 120-second R-band images over 2 -- 5 hours each night they observed (see Table \ref{tantab}).  Analysis of those images showed no significant trend or significant short-term brightness variation over the course of each observing session.  The photometry from individual images in the high cadence series were averaged together as single magnitude reported for the night.   

The brightness was measured using circular apertures adjusted for seeing conditions and sky background annuli around each aperture.  The aperture size and sky annuli were held constant when reducing a continuous series of images from a single telescope on the same night.  Twenty comparison stars within 10 arc minutes of the target were selected from the AAVSO Photometric All-Sky Survey\footnote{http://www.aavso.org/apass} \citep{2012JAVSO..40..430H}.  The Barber and Hambsch images were reduced using an unweighted average of results from all the comparison stars.  The Tan and Curtis images covered a smaller field and were reduced using an unweighted average of a sub-set of these comparisons.  A small correction was applied to Curtis' photometry to account for any systematic offsets introduced by the subset of comparisons used.   When pairs of images were taken by  Hambsch at roughly the same time, photometry was measured from the individual images and then averaged together.  
Statistical errors in the photometry for individual images was typically 0.05 magnitudes or less.  Photometry taken by different telescopes on the same night compare favorably with each other within the statistical errors.  Finally, the photometry was corrected to the system of \cite{2013ApJ...767....1P} using the corrections of:  dV = +0.063, dRc = +0.046 and dIc = +0.023 (Pastorello, personal communication).  There were no additional transformations for color or other factors applied to the measured magnitudes.

The Rc and Ic magnitudes are published in their entirety in \cite{2013arXiv1306.0038M}.  The V magnitudes prior to t = +10 days were previously published in \cite{2013ApJ...767....1P}.  The later V magnitudes are in Table \ref{vtab}.

\subsection{Swift UVOT Measurements}
SWIFT UVOT data in the b, u, w1, m1, and w2 bands are taken from \cite{2013arXiv1306.0038M} where they are described in detail.  The data were analyzed following the prescription of \cite{2009AJ....137.4517B} with apertures optimized to maximize signal-to-noise and limit contamination from nearby stars.  The magnitudes are on the UVOT photometric system \citep{2008MNRAS.383..627P} and revised to the Vega zero points of \citet{2011AIPC.1358..373B}.  The u and b data are converted to equivalent Johnson U and B magnitudes by \cite{2013arXiv1306.0038M}.  They note this is a minor correction that is {\em{not}\em} responsible for the bumps in the light curve.

\section{Analysis\label{analysis}}
\subsection{Optical Peak and Decline}
Our measurements begin on t = -9.16 days very near to the start of the rise in brightness toward the 2012-B peak \citep{2013ApJ...767....1P} .   The supernova brightness increased about 4 magnitudes in 12 days in V, Rc, and Ic, reaching a peak V = 13.80 $\pm$ 0.05 on t = +3 days (MJD 56206.04), which is consistent with observations reported by \cite{2013ApJ...763L..27P}.

The post peak bolometric flux of most Type IIn supernovae can be modeled by a power law function of flux with respect to time \citep{2012ApJ...759..108S,2013MNRAS.435.1520M}.  Bolometric magnitude is directly proportional to the logarithm of bolometric flux.  Therefore, a power law in flux-time is linear when transformed to a function of bolometric magnitude with respect to the logarithm of time.  For a sample whose duration spans a few decades of time, bolometric magnitude is nearly linear with respect to time.

The dominant trend in the decline in magnitude from the peak is almost linear with respect to time within the first 60 days (see Section \ref{dtrend}).  Fitting a purely linear function we obtained slopes of: 0.060 $\pm$ 0.005 mag/day in V, 0.050 $\pm$ 0.001 mag/day in Rc and 0.094 $\pm$ 0.026 mag/day in Ic.  The slopes differ due to different bolometric corrections  in each filter.  The rapid rise and rates of decline, within 60 days of the peak are consistent with other Type IIn light curves \citep{2012ApJ...744...10K,2013A&A...555A..10T}.

\subsection{Removal of the Underlying Trend\label{dtrend}}
The observed brightness after the peak contains significant fluctuations superimposed on a downward trend which are easily identified in all eight bands and the bolometric luminosity calculated by \cite{2013arXiv1306.0038M}.  Each telescope observed these fluctuations independently, eliminating the concern that it could be an effect attributable to a single telescope or observing location.  The record of the data presented here is also corroborated by the independent observations of \cite{2013MNRAS.433.1312F} and \cite{2014ApJ...787..163G}. \cite{2013arXiv1306.0038M} notes the regular placement of ``peaks'' in the general trend and that an analysis of the Rc-band photometry prior to the 2012-B event reveals some power in periods between 20 –- 40 days.  

\begin{figure}
\figurenum{1}
\label{loglog}
\epsscale{0.8}
\plotone{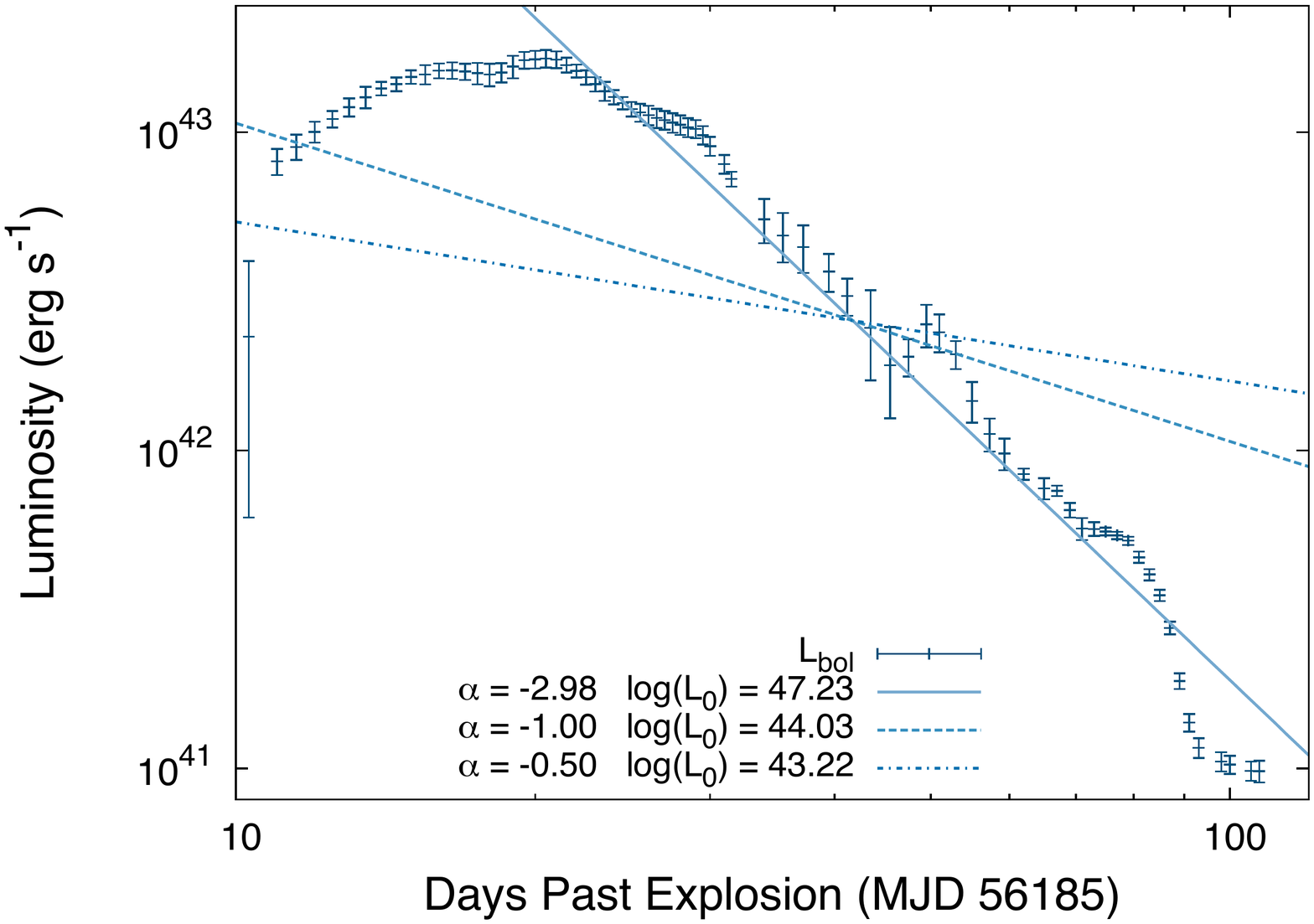}
\caption{A log-log plot of the bolometric luminosity from \cite{2013arXiv1306.0038M} fit by the Type IIn light curve model of \cite{2013MNRAS.435.1520M}.  A power law index of $\alpha$ = -2.98 is the best fit.  Other fits with slopes that are more consistent with the examples shown by \cite{2013MNRAS.435.1520M} are shown for comparison.}
\end{figure}

To analyze this structure, we subtracted the peak and linear decline from data in each band.  The best theoretical models of the trend after the peak are power law functions of bolometric flux with respect to time (eg.  \citet{2012ApJ...759..108S,2013MNRAS.435.1520M}) which are linearized as functions of bolometric magnitude with respect to the logarithm of time.  \cite{2013MNRAS.435.1520M} explicitly state that their model should not be applied when the circumstellar medium (CSM) is optically thick in the early stages of the Type IIn.  They also assume the power law governing the radial distribution of circumstellar material $s < 3$.  Figure \ref{loglog} shows fits of the Moriya's model to the bolometric luminosity curve published by \cite{2013arXiv1306.0038M}.  The power law of the best fit ($\alpha$ = -2.98) is much steeper than any of the examples given by \citet{2013MNRAS.435.1520M} (SN 2005ip, SN 2006jd, and SN 2010jl) and in all allowed self-similar cases for the density slope in the outer envelope of the progenitor yields a power law s = 5 for the circumstellar medium (CSM).  This result is outside the bounds of the model's assumptions and implies that the data are not consistent with the Type IIn model of \citet{2013MNRAS.435.1520M}.

\begin{figure}
\figurenum{2}
\label{figdetrend}
\plotone{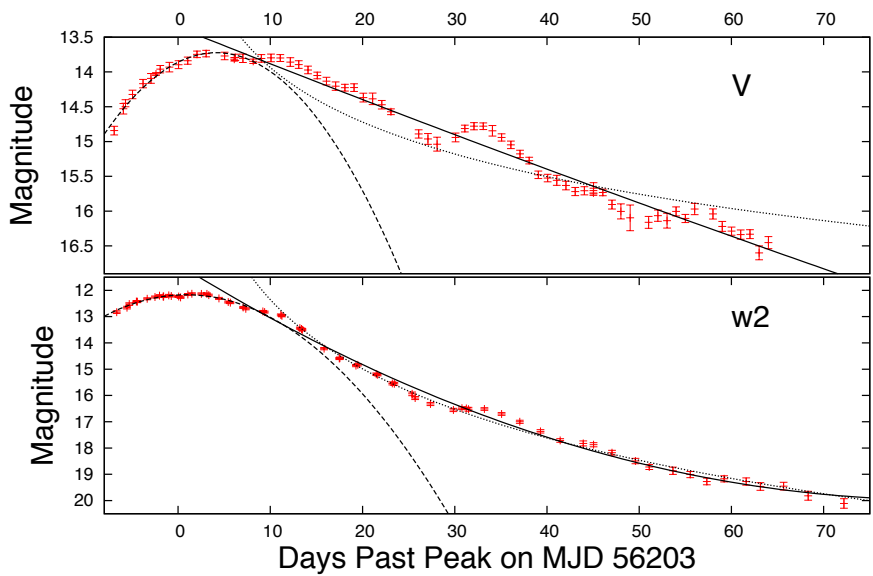}
\caption{An example of how the fits (see Table \ref{ptab}) were made to the V-band data (top panel) and w2 data (bottom panel).  The y-error bars are one sigma statistical errors.  The solid line and the dashed line are the second order polynomial fit between +10 $>$ t $>$ +75 days and -8 $<$ t  $<$ +10 days respectively.  The dotted line is the attempt at fitting a linear function with respect to Log(t) between +10 $>$ t $>$ +75 days.  The fit trends were subtracted from the data to produce Figure \ref{fig1}.}
\end{figure}

A more complicated model (i.e. the sum of two power laws in flux-time) is difficult to linearize and fit to the data using a least squares method.   A second degree polynomial of magnitude with respect to time is a reasonable low order approximation to the physical models.  \cite{2013arXiv1306.0038M} found significant transitions in the thermal continuum at t = +16 and t = +70 days.  The light curves in each band visibly change character at t $\sim$ +10 days and t $\sim$ +70 -- 75 days.  \cite{2014ApJ...787..163G} confirms a change in the color evolution of the light curve at the same times.  To accommodate this, our fit employed two second degree polynomials bracketed by these transitions (see example in Figure \ref{figdetrend}).  The first was fit to the data around the peak  for  -8 $<$ t  $<$ +10 days.  The second was fit to the decline between +10 $>$ t $>$ +75 days.  The coefficients for these fits are recorded in Table \ref{ptab}.  The small $\chi^2$ for these fits (Table \ref{ptab}) indicate that there is a high probability they are consistent with the trend in the data. 

For the purpose of comparison with our approach, we performed fits to the light curves in the interval of +10 $>$ t $>$ +75 days to a linear function of magnitude with respect to the logarithm of time.  The $\chi^2$ for all but one those fits (Table \ref{ptab}) demonstrate that they do not fit the data as well as the second degree polynomial.  In all cases, the function with respect to Log(t) is  a poor fit to the data at the extreme ends of the range (illustrated in Figure \ref{figdetrend}).  Note that the bolometric data were also better fit by our approach.  The one exception is the data for the w2 filter which is fit slightly better by the linear function with respect to Log(t) than the second degree polynomial.  

We acknowledge that there is a unquantifiable (but likely small) risk that  the second order polynomial over-subtracts the underlying trend.  A systematic influence of this type should increase red (Brownian) noise and reduce the significance of the bumps and dips.  If this effect is present there is also some risk that the relative heights of the peaks in different band-passes could be effected.  But it should have negligible effect on the spacing and frequency of the peaks.

The use of a relatively low order polynomial also risks under-subtraction of the underlying trend.  As with the risk of over-subtraction this may have increased the Brownian noise and possibly over emphasized the bumps in the decline.  To test this, similar fits were made to Type IIn supernova light curves from \cite{2013A&A...555A..10T}.  Subtraction of the fits generated using the same method from those light curves revealed no significant fluctuations at the one sigma level or greater relative to the quoted measurement errors in the individual data points.   If present, under-subtraction could also influence the relative heights of the peaks in each band-pass but should have a negligible effect on the spacing and frequency of the peaks.

\subsection{Brightness Fluctuations in the Decline\label{flux}}
The data with the modeled trends subtracted are plotted in Figure \ref{fig1}.   The detrended data has a series of peaks and troughs with an amplitude of 0.10 to 0.40 magnitudes in each band.  There are three larger dips followed by rises at approximately t = +5 -- +15 days, t = +20 -- +40 days, and t = +45 -- +65 days.  These fluctuations roughly match transition points in the evolution of the thermal continuum as modeled by \cite{2013arXiv1306.0038M} (see their Figure 8).   

There was no expectation of detecting changes on the time-scale of hours but Tan and Curtis were eager to test for the unexpected and/or provide a useful null result.  They made high cadence observations with a series of 120 second exposures spanning 2 -- 5 hours over 26 separate nights (in three circumstances both observing simultaneously on the same night, see Table \ref{tantab}).    A few of the observations (particularly the first observations by Tan during the steep rise prior to peak brightness) have a small but significant linear trend over the course of the night.  The observations revealed no significant fluctuations in the brightness of 2009ip on the time-scale of hours.  The standard deviations ($\sigma$) of measurements for each observing session are small.  There were a few instances where a series of continuous points fell more than two sigma from the average.  Careful inspection revealed these were caused by interference from an uncorrected hot pixel on the CCD.  Excepting those instances, there were no sudden flares or drops in the brightness of the target greater than $2\sigma$ from the mean over two or more consecutive measures.  

Figure \ref{VUm2} shows how the fluctuations evolve relative to each band over time.  All the bands start out fluctuating in phase and at roughly the same amplitude.  The bump at t $\sim$ +10 days is the first point where the bands separate, with the shorter wavelengths peaking at a higher amplitude.  At that point the UV bands (w1, w2, and m2) fall out of phase with the optical bands (Ic, Rc, V, B, and U).  The UV leads the optical into the dip at t $\sim$ +25 days and lags the optical by about one day coming out of the dip.   This lag was noted by \cite{2013MNRAS.433.1312F} but in contrast to their findings, we observe no lag in U relative to V and the amplitude measured from the low point at t $\sim$ +25 days to the peak at t $\sim$ +32 days differs significantly depending on wavelength.  At t $\sim$ +45 days the U and B bands fall out of phase with the V band, lagging that band by about 1 day and matching phase with the UV fluctuations.  The peak at t $\sim$ +60 days is brightest in the U band and slightly fainter in the B band, with the V and UV bands having roughly the same lower amplitude.  

These fluctuations are likely to be driven by continuum changes rather than by variations in the emission lines.  A change in flux of 0.1 magnitude in the filters used equals a change of 500$\AA$ -- 2000$\AA$ or more in equivalent width.   Adopting the distance modulus and bolometric correction for SN 2009ip used by \cite{2013arXiv1306.0038M}, the peaks represent an increase in total luminosity of approximately 10$^{42}$ erg s$^{-1}$.  The energy of a single peak (from the trend subtracted data) integrated over time is approximately 10$^{47}$ ergs.  No fluctuations matching these characteristics were reported in the spectrum of SN 2009ip  \citep{2013MNRAS.430.1801M,2013ApJ...767....1P,2013arXiv1306.0038M,2014ApJ...787..163G}.  

A peak or dip in brightness driven by a change in size of the continuum emitting region should affect the brightness in all filters equally.  But Figure \ref{VUm2} reveals {\em{relative}\em}  differences in color in each of the major peaks.   A relative heating of the continuum emitting region should cause shorter wavelengths to brighten relative to longer wavelengths while a relative cooling should cause longer wavelength to brighten relative to shorter wavelengths.  

In order to properly interpret the data, recall that we have subtracted the underlying trend so that differences in brightness between bands measure relative color.  An increase in brightness in shorter wavelength bands relative to longer wavelengths reveal the color becoming bluer, which can be interpreted as a wavelength dependent change in opacity (i.e. line blanketing) or relative heating.   Line blanketing can have the effect of reducing through increased opacity and enhancing through re-emission the brightness in specific filters leaving others unaffected.  Relative heating or cooling will proportionally effect all the filters according to the shift in peak of the blackbody continuum.  The analysis in \citet{2013arXiv1306.0038M} determined that while these fluctuations were occurring the continuum emitting region was expanding and cooling.   In the context of a black body continuum, a local peak in brightness can represent heating relative to the subtracted trend in declining temperature but the absolute temperature of each subsequent peak declines as time progresses.   Two peaks may have the same amplitude, revealing comparable heating, but the later one has a cooler absolute temperature.  

For the first peak at t $\sim$ +8 -- + 15 days and the following dip between t $\sim$ +20 -- +30 days the UV bands have a greater amplitude than the visual.  This could correspond to heating of the continuum emitting region to produce the peak and then cooling to produce the dip.  According to \cite{2013arXiv1306.0038M} this peak occurs when the effective temperature of the thermal continuum is between 15,000 K to 11,000 K.

The second major peak between t $\sim$ +30 -- +40 is different than the first.  All bands reach the same amplitude, but the UV bands lag the optical bands by $\sim$ 1 day.  This does not fit the expectation for a change driven by the shift in wavelength peak of a blackbody continuum. The lag in UV brightness is better understood in the context of line blanketing which simultaneously reduces brightness though increased opacity in one wavelength range and raises the brightness through re-emission in another.  Rising out of the minimum, significant line blanketing could simultaneously lower the UV flux and enhance the visual bands through re-emission.  This roughly corresponds to the explanation put forward by \cite{2013MNRAS.433.1312F} for the lag between bands in this peak. After the peak, the ionization of the continuum emitting region changes eliminating the line blanketing and causing the UV to be brighter throughout the decline.  According to \cite{2013arXiv1306.0038M} this peak occurs when the effective temperature of the thermal continuum is between 9500 K to 8000 K.

The third major peak shows a lag in both the UV and the blue optical bands.  This peak is brightest in the U and B bands and less prominent in the longer wavelength optical and UV.  The presence of the Balmer continuum/decrement in the U and B bands along with their divergent behavior relative to the other bands alludes to a transition in the ionization state of hydrogen.  According to \cite{2013arXiv1306.0038M} this peak occurs when the effective temperature of the thermal continuum is between 6500 K to 5500 K.  This peak occurs at a lower absolute temperature than the first two peaks.  There is probably some combination of line blanketing and temperature change at work.  Spectra from this time hint at an increase in structure at wavelengths shorter than 4000$\AA$  along with a significant increase in Balmer \citep{2013arXiv1306.0038M} and Paschen \citep{2013MNRAS.433.1312F} emission. 

\subsection{Power Spectrum Analysis of the Fluctuations}

The placement of the three major peaks and several smaller peaks imply a possible regularity in the fluctuations.  \cite{kashi} note the major bumps occur roughly 24 days apart and attributes this to binary interaction.  The behavior prior to t = +8 days is also very regular.  It should also be noted that the fluctuations after the eruption are consistent with those detected prior to the eruption as part of a binary or single star model.  Flickering (fluctuations on the order of 3 magnitudes in brightness over $\sim$ 16 days) was observed in earlier eruptions of SN 2009ip \citep{2010AJ....139.1451S,2011ApJ...732...32F}.   \cite{2013arXiv1306.0038M} also found periods of fluctuation between 20 -– 40 days in the R-band prior to the 2012-A eruption.   

There is already evidence for regularity in the fluctuations.  The power spectrum analysis that follows is performed to corroborate and quantify what is already supported by that evidence.  Without corroborating evidence, the results of our power spectrum analysis should be considered with caution. One objection may be that the data set is short and spans an insufficient length of time to yield meaningful results. To mitigate this concern we ignore the lowest frequencies in the power spectrum which correspond to periods more than one third the total time spanned by the data.  We also avoid tools like WOSA (Welch-Overlapped-Segment-Averaging, see \cite{1997Spectrum}) that subdivide the full data set into smaller segments for analysis.

Another concern is Brownian (red) noise. See the caveats and concerns outlined at the end of Section \ref{dtrend}.  The process to subtract the underlying trend is certain to introduce noise heavily weighted toward low frequencies. To answer those concerns we have applied a tested and peer-reviewed algorithm to estimate the confidence level of the peaks relative to Brownian noise (see below).  A third concern is white noise unequally affecting individual data points.  To address this we have modeled their influence with a Monte Carlo simulation.  There may still be doubts about the validity of the power spectrum analysis results on their own, but to the extent that they corroborate other observations they are compelling.

\begin{figure}
\figurenum{3}
\label{fig1}
\epsscale{0.8}
\plotone{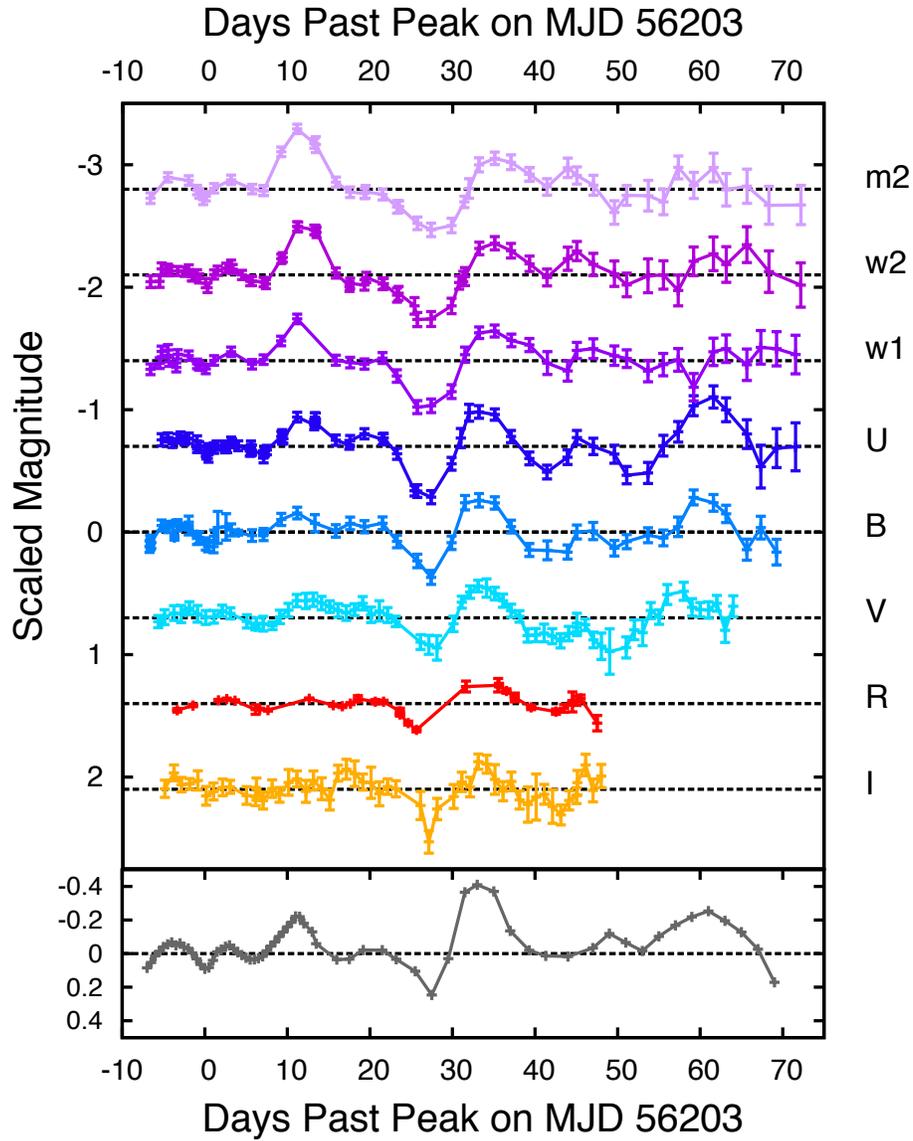}
\caption{Brightness vs time during 2012-B outburst with the fit to the decline subtracted.  Magnitudes measured in each filter are shifted relative to each other.  Error bars are one sigma.  The bottom plot is the bolometric magnitude from \cite{2013arXiv1306.0038M} with the decline subtracted.}
\end{figure}

\begin{figure}
\figurenum{4}
\label{VUm2}
\plotone{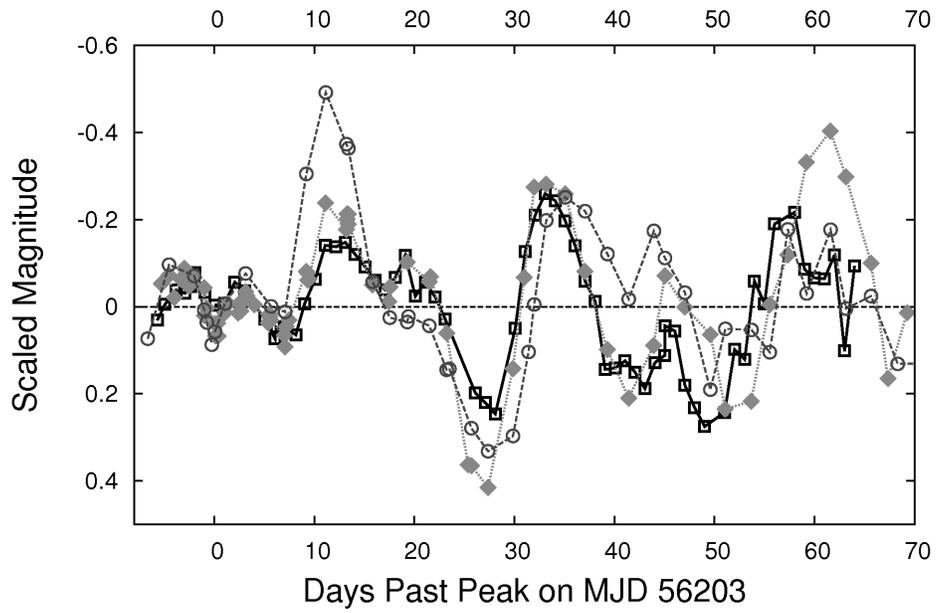}
\caption{Detrended magnitudes for the V (dark open squares + solid line), U (gray shaded diamonds + dotted line), and m2 (open circles + dashed line) plotted on the same scale to highlight the differing amplitudes and relative lag of the fluctuations as a function of wavelength.}
\end{figure}

We analyzed the fluctuations using a date compensated discrete Fourier transform (DCDFT) \citep{1981AJ.....86..619F} in the AAVSO software package VStar version 2.16.1\footnote{http://www.aavso.org/vstar-overview}.  The analysis was performed on the detrended data (Figure \ref{fig1}).  We will focus primarily on the power spectrum of the bolometric magnitude (bottom panel in Figure \ref{fig1}).  That spectrum has notable peaks at 0.043 $day^{-1}$ (period $\sim$ 23-days), 0.079 $day^{-1}$ (period $\sim$ 13-days), and 0.132 $day^{-1}$ (period $\sim$ 8-days).  The frequencies of the peaks are almost integer multiples of each other.  There are also hints of weaker adjacent peaks at 0.058 $day^{-1}$, 0.095 $day^{-1}$ and 0.145 $day^{-1}$.   The most significant peaks in the M$_{bol}$ power spectrum are in the power spectra for each individual photometric band and the power spectrum of the bolometric magnitude data {\em{without}\em} the trend subtracted (dashed line, bottom panel).  V, Rc, and Ic data from  \cite{2013MNRAS.433.1312F} processed in the same way independently support these results.

The process by which the underlying trend was subtracted is likely to introduce Brownian noise into the data.  This is evident in Figure \ref{fig1} as a continuous decrease in spectral amplitude with increasing frequency.  Similar noise dominates plaeoclimate data \citep{1976Hasselmann} and can be modeled using a first-order autoregressive (AR1) process.  \citet{2002Redfit} provide an algorithm to model this noise in  an unevenly spaced time series and simulate confidence levels in order to determine the significance of peaks in the power spectrum.  Figure \ref{rednoise} shows the results of applying their code to our data using N = $10^{4}$ for the Monte-Carlo simulation.  Three major peaks and one weaker adjacent peak are above the 99\% confidence level in the REDFIT simulation.  Peaks above a 99\% confidence level have less than a 1\% chance of being a false alarm.  If our sample is dominated by Brownian noise this simulation implies that several of the observed peaks in the power spectrum are {\em{not}\em} a product of AR1-type Brownian noise.

However, it is unclear to what extent Brownian noise may {\em{dominate}\em} this data (a key assumption of the REDFIT analysis).  The white noise from photometric errors could play a more significant role.  Irregular spacing of the points in time combined with white noise has a non-linear influence on the power spectrum.  We modeled this with a Monte Carlo simulation of one thousand perturbed data sets generated and run through the same REDFIT analysis.  In each simulated set the bolometric magnitude was perturbed by an amount randomly selected from a normal distribution with a standard deviation equal to the photometric error.  The times were perturbed in a similar manner with a standard deviation of 0.05 days\footnote{A time error with a standard deviation of $\pm$0.05 days is large and unlikely.  The longest exposures for our data are on the order of ten minutes (0.007 days).  Our simulation showed that the outcome was mostly insensitive to errors in time smaller than 0.1 days.}.  

The results of the Monte Carlo simulation are presented in Figure \ref{noisesim}.  The the peaks for average outcome in the simulation (dashed white line) are at a lower confidence level than the results presented in Figure \ref{fig02}.  This indicates that the white noise has a greater influence on the power spectrum than the Brownian noise.   In more than half of the simulated power spectra the three highest peaks have a confidence level of 80\% or greater (false alarm rate of 20\% or less).  As a result of this simulation we have confidence in the first two peaks.  In more than 99\% of the simulated power spectra the first peak (0.043 $day^{-1}$) was higher than the modeled AR1 noise level.  More than 75\% of the simulated power spectra had a REDFIT false alarm rate less than 20\% for this peak.   The second peak (0.079 $day^{-1}$) was above the modeled AR1 noise level in more than 95\% of the simulations and more than 80\% of the simulated power spectra had a REDFIT false alarm rate less than 20\% for this peak.  Conservatively speaking, this implies a better than 1.5 sigma detection for both these peaks. Combined with our other findings, it appears likely that the first two peaks (and likely more) are an actual signal and {\em{not}\em} a false alarm.  Our confidence in several of the less significant peaks is bolstered by their status as sub-harmonics of more significant frequency peaks.

\section{Discussion \label{discuss}}
The light curve for the 2012-B outburst was exceptionally well-sampled raising the possibility that these fluctuations are common in the light-curves of Type IIn and have only been uncovered as a result of greater scrutiny.  While the 2012-B eruption of SN 2009ip clearly meets the spectroscopic definition of a Type IIn \citep{2012ATel.4412....1S,2013MNRAS.430.1801M,2013ApJ...767....1P,2013MNRAS.433.1312F,2013arXiv1306.0038M,2013arXiv1308.0112S}, and the peak absolute magnitude is consistent with a terminal core collapse \citep{2013MNRAS.430.1801M,2013ApJ...767....1P,2013MNRAS.433.1312F,2013arXiv1306.0038M}, the bumpy decline of SN 2009ip is not prevalent among other Type IIn’s  \citep{2012ApJ...744...10K,2013A&A...555A..10T}.  \cite{2013MNRAS.433.1312F} note that the fluctuations in SN 2009ip share some characteristics with SN 2005la which \cite{2008MNRAS.389..131P} classified type-Ibn (sharing some characteristics of ``classical'' Type IIn events).  The light curves of a few other Type IIn events (i.e SN 1994W \citep{1998ApJ...493..933S}, SN 2005la \citep{2008MNRAS.389..131P}, SN 2005ip \citep{2009ApJ...695.1334S} and SN 2010mc \citep{2013Natur.494...65O}) appear to have similar fluctuations however they have never been discussed in length.  

Interaction of the expanding supernova shock-wave with circumstellar material or other related mechanisms could cause fluctuations in brightness after the peak of the light curve.  Many of the models which are proposed for the 2012-B eruption of SN 2009ip employ this mechanism (see Section \ref{intro}).  \cite{2014ApJ...787..163G} demonstrate some correlation between the past activity of SN 2009ip and the bumps in the light curve.  However, none those models clearly and quantitatively address the observed fluctuations.  Some permutation of those models may be capable of reproducing the fluctuations under tuned conditions.  It remains to be proven if those conditions are reasonable or not.  Further exploration along those lines is outside the scope of this paper and best left to others with access to the appropriate expertise and modeling codes.  

Type-Ia supernovae sometimes show a secondary peak powered by continuum brightening in the red and infrared spectral region \citep{1981ApJ...251L..13E}. The origin of the second Ia peak is reprocessed radiation driven by recombination and subsequent line blanketing in the ultraviolet part of the spectrum \citep{2006ApJ...649..939K}.  We note a correlation between a change in ionization state/opacity in the thermal continuum blackbody fits of \cite{2013arXiv1306.0038M} and the second and third major peaks (Section \ref{flux}).  \cite{2013MNRAS.433.1312F} postulate that scattering could play a role in the dips and peaks.  However, unlike the secondary peaks in type-Ia, the fluctuations in the decline of SN 2009ip were {\em{less}\em} prominent in the red and NIR part of the spectrum \citep{2013arXiv1306.0038M}.    

Under specially constructed circumstances the light curve fluctuations can be modeled with light echoes.  It is possible to contrive an environment where they are reproduced by shells of the proper shape, orientation, and solid angle to introduce the bumps as scattered light echos of the primary eruption.  The periodic nature of the bumps could be explained by nested shells expelled during periodic episodes of mass loss from the progenitor.  It is common to find nested shells at some regular frequency around stars at late stages in their evolution.  However, the albedo for the scattering and/or the effective solid angle subtended by these structures would have to be exceptionally high in order to produce echos on the order of several tens of percent of the brightness.  The geometric tolerances are further constrained by the delays of the major peaks.  Delays of +10, +30 and +60 days from the peak would require distances of 870 AU, 2600 AU, and 5200 AU respectively between the progenitor and the near-point on the semi-major axis of properly oriented partial ellipsoidal shells.  The model would also have to successfully explain the key differences in brightness as a function of filter in each of the successive peaks described in section \ref{flux}.  We cannot rule out light echos as a possible explanation for the bumps but it appears to require a very complicated model to satisfy all our observations.

\begin{figure}
\figurenum{5}
\label{fig02}
\epsscale{0.85}
\plotone{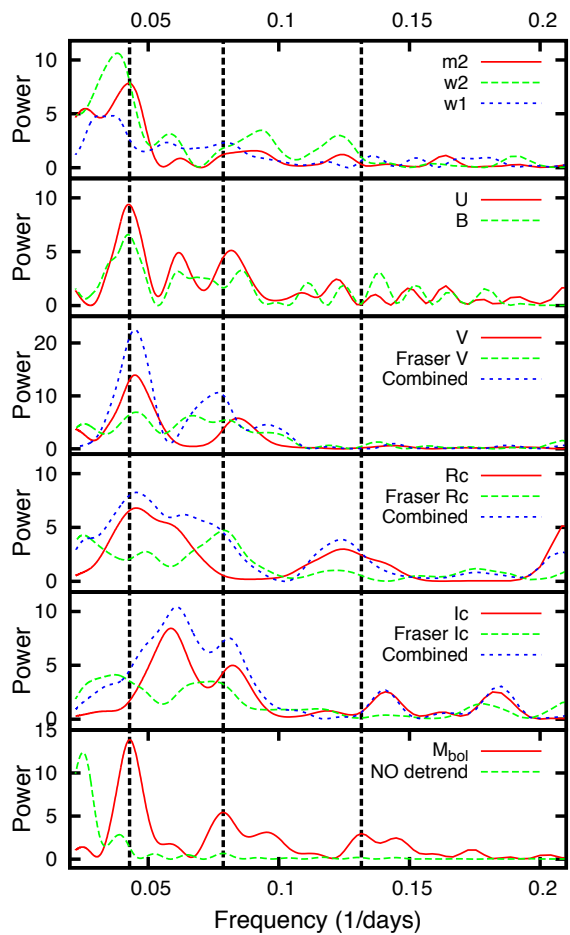}
\caption{Power spectra derived for the detrended data.  The dashed line in the bottom panel is the power spectrum of the bolometric magnitude data without the 
trend subtracted.   The results from the same analysis of independent data from \cite{2013MNRAS.433.1312F} are included for comparison with the V, Rc, and Ic bands.  The vertical dashed lines mark the peaks of note in the power spectrum of the bolometric magnitude (bottom panel) at 24 days, 12 days, and 8 days.}
\end{figure}

\begin{figure}
\figurenum{6}
\label{rednoise}
\plotone{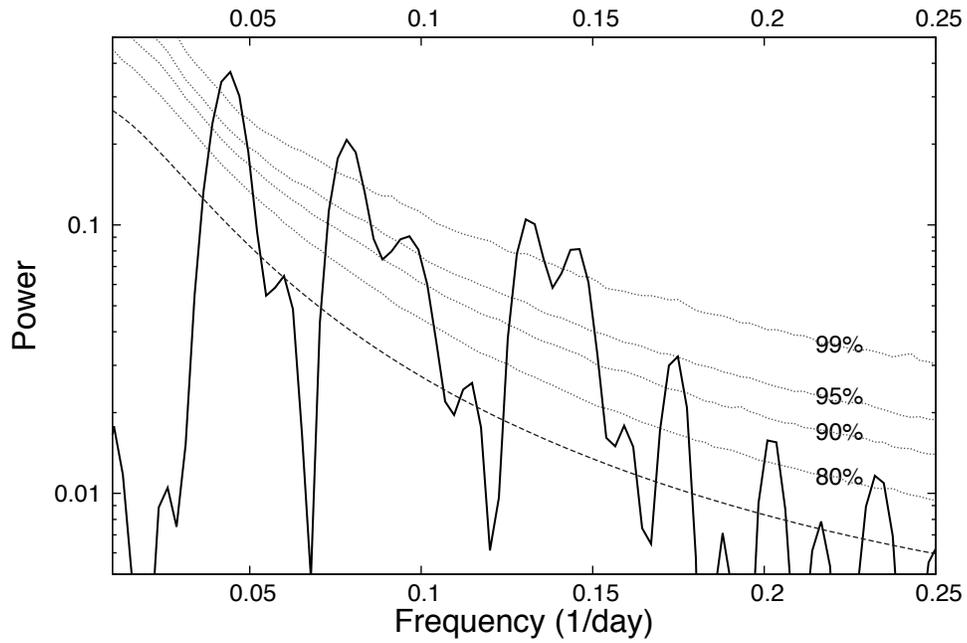}
\caption{A power spectrum of the the detrended bolometric magnitude data compared with an AR1 red-noise simulation \citep{2002Redfit} (dashed line).  The dotted lines depict the increasing levels of confidence calculated using a Monte-Carlo simulation with N = $10^{4}$.  Spectrum above the 95\% confidence level has a less than 5\% chance of being a false alarm.}
\end{figure}

\begin{figure}
\figurenum{7}
\label{noisesim}
\plotone{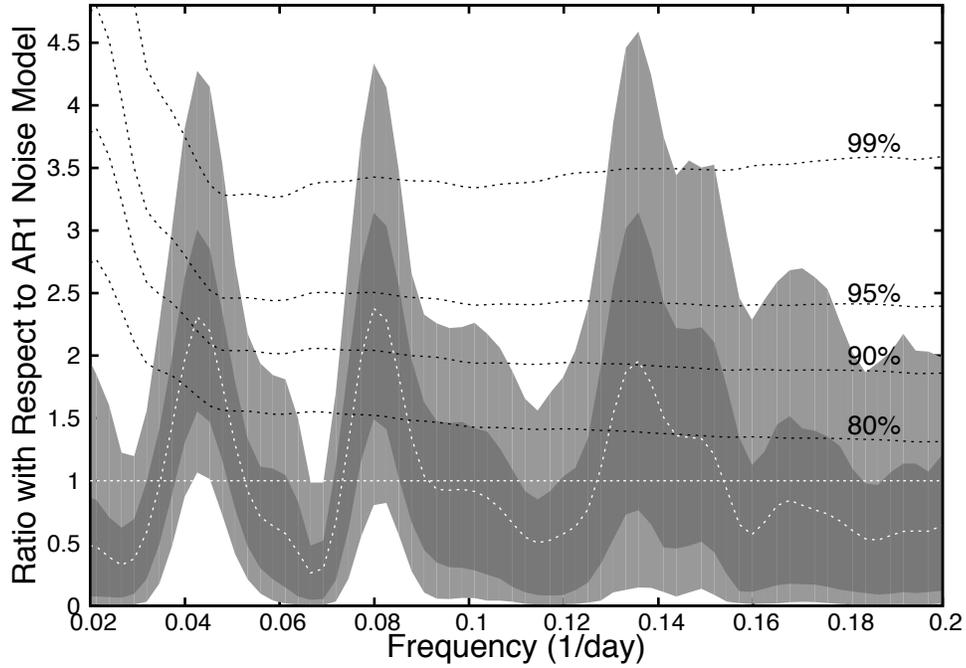}
\caption{A plot of the ratio of the power spectrum with respect to the AR1 Brownian noise model.  The results of the Monte Carlo simulation of white noise are plotted as follows:  the average of the simulations (white dashed line), the area containing simulations falling with in one sigma of the mean simulation (dark gray shading), and the area containing simulations falling between one and two sigma from the mean (light gray shading).  The confidence intervals calculated by REDFIT are plotted as dashed black lines. }
\end{figure}

\cite{2013ApJ...764L...6S} and \cite{kashi} model the fluctuations as interactions between shells of material excreted from a binary system embedded inside a slower moving cocoon of material.  The power spectrum (Figure \ref{fig02}, \ref{rednoise}, and \ref{noisesim}) with frequency peaks and sub-harmonics could be generated by a periodic system.  There is also a slight similarity with the power spectrum of pulsations in a single star (the amplitude of the brightness fluctuations are probably too large to comfortably explain in this way).  The purpose of this analysis is not to say conclusively that a pulsating progenitor survived the explosive event and contributed to the brightness fluctuations during the decline although if it were like Eta Carinae's Great Eruption, it may have.  \cite{2013arXiv1306.0038M} found some evidence of similar periodicity in the photometry before the 2012-B eruption.  The pre-event activity of the progenitor must have sculpted the circumstellar medium.  In some models (eg. \cite{2014ApJ...787..163G}) the pre-event activity of the star influences the brightness fluctuations through the regular or irregular placement of inhomogeneities in the CSM.  So an analysis of these frequencies as adiabatic pulsations of a single star may also independently corroborate work to determine the state of the progenitor.

Following the prescription of \cite{1926ics..book.....E} we computed the mean density of a star ($\overline{\rho}$) undergoing adiabatic radial pulsation at certain frequencies (f) using the relation:

$\overline{\rho} = \pi {\left[ \frac{f}{(\gamma - \frac{4}{3})G} \right]}^{2}$

Assuming the effective ratio of specific heats for the pressure-density relation is $\gamma = \frac{5}{3}$ and $G = 6.67 \times 10^{-8} cm^{3} g^{-1} s^{-1}$.  The resulting mean stellar density, log($^{\bar{\rho}}/_{\bar{\rho}_\sun}$) $\sim$ -4.0 to -4.6 for frequencies of 8 to 24 days, consistent with an A-type super-giant with a surface temperature between 10,000 K and 8000 K and an absolute visual magnitude between -8.5 and -9.0 or a late-K type giant with a surface temperature between 4,000K and 3800K and an absolute visual magnitude between 0.0 and -0.4 \citep{1992adps.book.....L}.  An  A-type super-giant is consistent with the proposed progenitor of SN 2009ip positioned near the Humphrey-Davidson limit on the HR-diagram \citep{2010AJ....139.1451S,2011ApJ...732...32F}.  This is an interesting coincidence.  There are many additional details that must be considered to incorporate this information into a viable model.   

The models for Type IIn SNe have encountered a number of paradoxes in SN 2009ip.  Although their nature remains uncertain, the bumps in the decline from the 2012-B eruption appear to be real and there is further evidence that they are periodic.  This phenomena deserves further study.  Do these fluctuations in some Type IIn have a common origin?  Perhaps they can provide clues as to the nature of this class of supernovae.

\acknowledgments
J.C. Martin's work is supported by the National Science Foundation grant AST–1108890.
R. Margutti thanks the Swift team for their excellent support in scheduling the observations.
We wish also to thank  Jim O'Brien and Jennifer Hubbell-Thomas for assisting with observations included in this publication and the patrons of the UIS Barber Observatory for their financial support of the Henry R. Barber endowment for Astronomy at the University of Illinois Springfield.
Thanks to R.M. Humphreys and K. Davidson who provided several constructive comments.  Thank you also to the anonymous referees whose thoughtful criticism greatly improved this paper.
J.C.\ Martin makes a heartfelt acknowledgment of contributions by George W. Collins III made posthumously through notations recorded in the margin of Martin's copy of \cite{1926ics..book.....E}.


{\it Facilities:} \facility{Swift}.

\clearpage


\clearpage

\begin{deluxetable}{lccccccc}
\tablewidth{0pt}
\tabletypesize{\tiny}
\tablecolumns{8}
\tablecaption{Record of High Cadence Observations by Tan \& Curtis\label{tantab}}
\tablehead{
\colhead{Date}&
&
\colhead{Average}&
\colhead{$\sigma$}&
\colhead{Number of}&
\colhead{Time Span}&
\colhead{Fit Slope}&
\\
\colhead{(MJD)}&
\colhead{Filter}&
\colhead{(mag)}&
\colhead{(mag)}&
\colhead{Exposures\tablenotemark{a}}&
\colhead{(hours)}&
\colhead{(mag/day)}&
\colhead{Source}
}
\startdata 
56199.6&Rc&14.010&0.014&140&5.5&-0.128$\pm$0.014&Tan\\
56201.5&Rc&13.864&0.010&80&3.1&-0.062$\pm$0.029&Tan\\
56204.6&Rc&13.696&0.007&113&4.3&-0.049$\pm$0.013&Tan\\
56205.6&Rc&13.659&0.008&139&5.3&-0.032$\pm$0.010&Tan\\
56206.6&Rc&13.652&0.007&56&5.3&-0.038$\pm$0.012&Tan\\
56209.6&Rc&13.696&0.007&131&5.0&0.012$\pm$0.010&Tan\\
56210.6&Rc&13.711&0.008&128&4.9&-0.023$\pm$0.011&Tan\\
56215.6&Rc&13.750&0.006&67&2.6&0.027$\pm$0.022&Tan\\
56218.5&Rc&13.947&0.009&51&2.4&0.115$\pm$0.041&Tan\\
56219.6&Rc&14.010&0.010&106&4.2&0.009$\pm$0.022&Tan\\
56220.6&Rc&14.038&0.008&110&4.2&0.018$\pm$0.015&Tan\\
56221.5&Rc&14.043&0.025&210&4.7&0.033$\pm$0.025&Curtis\\
56223.6&Rc&14.171&0.017&99&3.9&0.053$\pm$0.035&Tan\\
56224.6&Rc&14.221&0.012&91&3.8&0.053$\pm$0.025&Tan\\
56226.6&Rc&14.425&0.013&97&3.8&0.068$\pm$0.029&Tan\\
56226.6&Rc&14.399&0.026&141&3.7&0.127$\pm$0.049&Curtis\\
56227.6&Rc&14.544&0.016&92&3.6&0.021$\pm$0.038&Tan\\
56228.6&Rc&14.649&0.017&79&3.0&0.123$\pm$0.051&Tan\\
56238.6&Rc&14.780&0.015&66&2.6&0.083$\pm$0.057&Tan\\
56239.5&Rc&14.873&0.018&50&2.0&-0.018$\pm$0.112&Tan\\
56240.5&Rc&14.975&0.036&90&3.2&0.067$\pm$0.100&Curtis\\
56240.5&Rc&14.976&0.019&66&2.5&0.083$\pm$0.057&Tan\\
56242.5&Rc&15.155&0.019&49&2.4&0.076$\pm$0.087&Tan\\
56245.5&Rc&15.339&0.022&55&2.1&-0.233$\pm$0.114&Tan\\
56246.5&Rc&15.362&0.031&54&2.0&0.210$\pm$0.169&Tan\\
56247.5&Rc&15.359&0.081&47&1.6&-0.029$\pm$0.614&Curtis\\
56247.5&Rc&15.366&0.019&50&1.9&-0.370$\pm$0.114&Tan\\
56248.5&Rc&15.380&0.028&46&1.8&-0.364$\pm$0.186&Tan\\
56250.5&Rc&15.681&0.063&32&1.1&0.066$\pm$0.610&Curtis\\
\enddata
\tablenotetext{a}{All exposures were 120 seconds long.}
\end{deluxetable}

\begin{deluxetable}{lcc}
\tablewidth{0pt}
\tabletypesize{\tiny}
\tablecolumns{3}
\tablecaption{V Magnitudes After t = +10 days\label{vtab}}
\tablehead{
\colhead{Date}&
\colhead{V}&
\\
\colhead{(MJD)}&
\colhead{(mag)}&
\colhead{Telescope}
}
\startdata 
56213.08&13.798$\pm$0.055&Hambsch\\
56214.08&13.800$\pm$0.050&Hambsch\\
56215.17&13.858$\pm$0.062&Hambsch\\
56216.09&13.894$\pm$0.055&Hambsch\\
56217.09&13.972$\pm$0.055&Hambsch\\
56218.09&14.051$\pm$0.048&Hambsch\\
56219.09&14.132$\pm$0.055&Hambsch\\
56220.09&14.204$\pm$0.060&Hambsch\\
56221.09&14.226$\pm$0.048&Hambsch\\
56222.09&14.225$\pm$0.055&Hambsch\\
56223.09&14.369$\pm$0.063&Hambsch\\
56224.09&14.387$\pm$0.083&Hambsch\\
56225.09&14.471$\pm$0.061&Hambsch\\
56226.09&14.572$\pm$0.042&Hambsch\\
56229.09&14.891$\pm$0.060&Hambsch\\
56230.09&14.963$\pm$0.077&Hambsch\\
56231.09&15.038$\pm$0.100&Hambsch\\
56233.09&14.941$\pm$0.063&Hambsch\\
56234.09&14.814$\pm$0.049&Hambsch\\
56235.09&14.780$\pm$0.051&Hambsch\\
56236.09&14.780$\pm$0.048&Hambsch\\
56237.08&14.846$\pm$0.075&Hambsch\\
56238.08&14.941$\pm$0.052&Hambsch\\
56239.08&15.048$\pm$0.054&Hambsch\\
56240.08&15.178$\pm$0.053&Hambsch\\
56241.07&15.274$\pm$0.047&Hambsch\\
56242.07&15.479$\pm$0.054&Hambsch\\
56243.07&15.525$\pm$0.054&Hambsch\\
56244.07&15.557$\pm$0.072&Hambsch\\
56245.06&15.632$\pm$0.062&Hambsch\\
56246.06&15.719$\pm$0.058&Hambsch\\
56247.06&15.708$\pm$0.059&Hambsch\\
56248.04&15.739$\pm$0.034&Barber\\
56248.05&15.672$\pm$0.081&Hambsch\\
56249.05&15.733$\pm$0.047&Hambsch\\
56250.05&15.906$\pm$0.064&Hambsch\\
56251.02&16.005$\pm$0.108&Hambsch\\
56252.02&16.097$\pm$0.185&Hambsch\\
56254.02&16.162$\pm$0.086&Hambsch\\
56255.02&16.066$\pm$0.081&Hambsch\\
56256.02&16.138$\pm$0.106&Hambsch\\
56257.00&16.005$\pm$0.065&Hambsch\\
56258.00&16.106$\pm$0.059&Hambsch\\
56259.01&15.971$\pm$0.081&Hambsch\\
56261.00&16.041$\pm$0.073&Hambsch\\
56262.01&16.220$\pm$0.071&Hambsch\\
56263.01&16.288$\pm$0.058&Hambsch\\
56264.00&16.338$\pm$0.067&Hambsch\\
56265.01&16.332$\pm$0.064&Hambsch\\
56266.01&16.600$\pm$0.100&Hambsch\\
56267.01&16.453$\pm$0.086&Hambsch\\
\enddata
\end{deluxetable}

\begin{deluxetable}{lcccccccccccccc}
\rotate
\tablewidth{0pt}
\tabletypesize{\tiny}
\tablecolumns{15}
\tablecaption{Parameters For Light Curve Fits\label{ptab}}
\tablehead{
&
\colhead{a$_1$ (mag)}&
\colhead{b$_1 \times 10^2$}&
\colhead{c$_1 \times 10^3$}&
&
\colhead{$(\chi^2)/N$}&
\colhead{a$_2$ (mag)}&
\colhead{b$_2 \times 10^2$}&
\colhead{c$_2 \times 10^3$}&
&
\colhead{$(\chi^2)/N$}&
\colhead{a$_3$ (mag)}&
\colhead{b$_3 \times 10^2$}&
&
\colhead{$(\chi^2)/N$}
\\
\colhead{Filter}&
\colhead{(mag)}&
\colhead{(mag/day)}&
\colhead{(mag/day$^2$)}&
\colhead{N}&
\colhead{x100}&
\colhead{(mag)}&
\colhead{(mag/day)}&
\colhead{(mag/day$^2$)}&
\colhead{N}&
\colhead{x100}&
\colhead{(mag)}&
\colhead{(mag/Log(day))}&
\colhead{N}&
\colhead{x100}
}
\startdata 
Ic&13.81$\pm$0.02&-5.5$\pm$0.5&3.5$\pm$0.5&18&0.48&13.02$\pm$0.14&5.5$\pm$1.1&\tablenotemark{a}&34&1.54&10.71$\pm$0.17&1.14$\pm$0.05&35&2.12\\
Rc&13.76$\pm$0.01&-5.2$\pm$0.2&6.2$\pm$0.4&5&0.02&13.22$\pm$0.05&4.8$\pm$0.2&\tablenotemark{a}&24&0.92&10.80$\pm$0.26&1.17$\pm$0.08&24&3.43\\
V&13.86$\pm$0.02&-6.6$\pm$0.3&7.9$\pm$0.7&16&0.35&13.36$\pm$0.07&5.3$\pm$0.5&\tablenotemark{a}&55&2.07&11.33$\pm$0.21&1.13$\pm$0.06&56&12.07\\
B&13.90$\pm$0.02&-6.4$\pm$0.3&8.7$\pm$0.7&25&0.52&13.62$\pm$0.19&4.5$\pm$1.0&0.3$\pm$0.1&29&3.96&7.07$\pm$0.58&2.53$\pm$0.16&30&21.19\\
U&12.91$\pm$0.02&-5.5$\pm$0.3&8.1$\pm$0.5&35&0.45&12.15$\pm$0.15&9.3$\pm$0.9&\tablenotemark{a}&37&4.41&5.10$\pm$0.43&3.03$\pm$0.12&38&19.40\\
w1&12.25$\pm$0.02&-3.3$\pm$0.4&9.5$\pm$0.8&13&0.39&11.48$\pm$0.17&15.0$\pm$0.9&-0.6$\pm$0.1&28&2.61&3.70$\pm$0.37&3.53$\pm$0.10&29&7.29\\
m2&12.09$\pm$0.03&-2.7$\pm$0.6&9.6$\pm$1.0&9&0.60&10.97$\pm$0.13&19.6$\pm$0.6&-1.0$\pm$0.1&35&3.39&3.05$\pm$0.24&3.86$\pm$0.07&36&5.03\\
w2&12.18$\pm$0.02&-1.8$\pm$0.3&10.3$\pm$0.6&23&0.38&11.01$\pm$0.15&21.7$\pm$0.9&-1.3$\pm$0.1&37&3.92&3.62$\pm$0.17&3.80$\pm$0.05&38&2.77\\
M$_{bol}$&-19.34$\pm$0.01&-2.5$\pm$0.2&10.5$\pm$0.5&28&0.23&-19.70$\pm$0.08&10.6$\pm$0.5&-0.6$\pm$0.1&37&1.99&-24.07$\pm$0.28&2.11$\pm$0.08&35&98.08\\
\enddata
\tablenotetext{a}{The fit for this term was consistent with zero.}
\tablecomments{These are the best fit parameters to the measured brightness in each band for the time periods {-8 $<$ t $<$ +10 days} (brightness = {a$_1$+ b$_1$t+c$_1$t$^2$}) and {+10 $<$ t $<$ 75 days} (brightness = {a$_2$+ b$_2$t+c$_2$t$^2$)}.  For comparison, the fit parameters and $\chi^2$ for {brightness = a$_3$+b$_3$Log(t)} to {+10 $<$ t $<$ 75 days} are also included. }

\end{deluxetable}

\end{document}